\documentclass{chi-ext}

\title{Can Multisensory Cues in VR Help Train Pattern Recognition to Citizen Scientists?}

\numberofauthors{1}
\author{
  \alignauthor{
  	\textbf{Alina Striner}\\
  	\affaddr{School of Information Science }\\
  	\affaddr{College Park, MD 20742 USA}\\
  	\email{algol001@umd.edu}
  }
  \vfil
}

\def\plaintitle{CHI LaTeX Extended Abstracts Template}
\def\plainauthor{Luis A. Leiva}
\def\plainkeywords{Qualitative Judgments; Citizen Science; Collaborative Learning;}
\def\plaingeneralterms{Virtual worlds; training simulations; Software Design techniques; User Centered Design}

\hypersetup{
  pdftitle={\plaintitle},
  pdfauthor={\plainauthor},  
  pdfkeywords={\plainkeywords},
  pdfsubject={\plaingeneralterms},
}

\usepackage{graphicx}   
\usepackage{balance}    
\usepackage{bibspacing} 
\usepackage{marginnote}
\usepackage{textcomp} 

\usepackage{array}
\newcolumntype{L}[1]{>{\raggedright\let\newline\\\arraybackslash\hspace{0pt}}m{#1}}

\begin{document}

\maketitle

\begin{abstract}
As the internet of things (IoT) has integrated physical and digital technologies, designing for multiple sensory media ("mulsemedia") has become more attainable. Designing technology for multiple senses has the capacity to improve virtual realism, extend our ability to process information, and more easily transfer knowledge between physical and digital environments.  HCI researchers are beginning to explore the viability of integrating multimedia into virtual experiences, however research has yet to consider whether mulsemedia truly enhances pattern recognition and knowledge transfer in virtual reality (VR). My work on StreamBED, a VR training to help citizen scientists make qualitative judgments of stream environments, plans to consider the role of mulsemedia in observation and pattern recognition. Future findings about the role of mulsemedia in learning contexts will potentially allow learners to experience, connect to, learn from spaces that are impossible to experience firsthand.
\end{abstract}


\category{H.5.m}{{Virtual worlds} {training simulations} {Software Design techniques} {User Centered Design}}.

\section{Introduction}

Since the early days of Star Trek, VR researchers have pursued the realism of the Holodeck (figure \ref{fig:Holodeck}), a fully immersive experience with the potential to virtually recreate the physical world using the five primary senses, visual (sight), auditory (sound), tactile/haptic (touch), olfactory (smell), and gustatory (taste)~\cite{ghinea2014mulsemedia}. Heilig patented the multisensory Sensorama machine as early as 1962~\cite{gallace2012multisensory} (figure~\ref{fig:Sensorama}), however only recently have researchers like Jacob~\cite{jacob2008reality} have  suggested the shift of technology away from disembodied interactions toward ``reality-based'' experiences that build upon our existing knowledge of the world. As IOT technologies have become more integrated into end user experience, designing for multiple senses has transformed from a TV fantasy into a feasible goal. 

In contrast to traditional media comprised of audio and video senses, Ghinea~\cite{ghinea2014mulsemedia} defines "mulsemedia" as media  that incorporates three or more senses. Research suggests that mulsemedia has many benefits, allowing users to more easily process and interpret information. Ghinea, for instance, describes how sensory information is processed and stored differently by the brain; unlike visual and auditory information, tacitle, olfactory and gustatory information contributes to episodic knowledges that helps shape attention. Likewise, Haverkamp~\cite{haverkamp2001application} suggests that communicating information through multiple sensory channels allow for optimal information processing on multiple levels of consciousness.

Literature also suggests that sensory modalities affect one another; Krishna et. al.~\cite{krishna2010feminine} describe the presence of multisensory congruence, showing that sensory experience in different modalities (e.g. touch and smell) 
can impact one another, and Fujisaki~\cite{fujisaki2015perception} and Donley~\cite{donley2014_MultisensoryQOE} even suggest the 
\reversemarginpar
\marginnote{
\begin{figure}
\includegraphics[width=\marginparwidth]{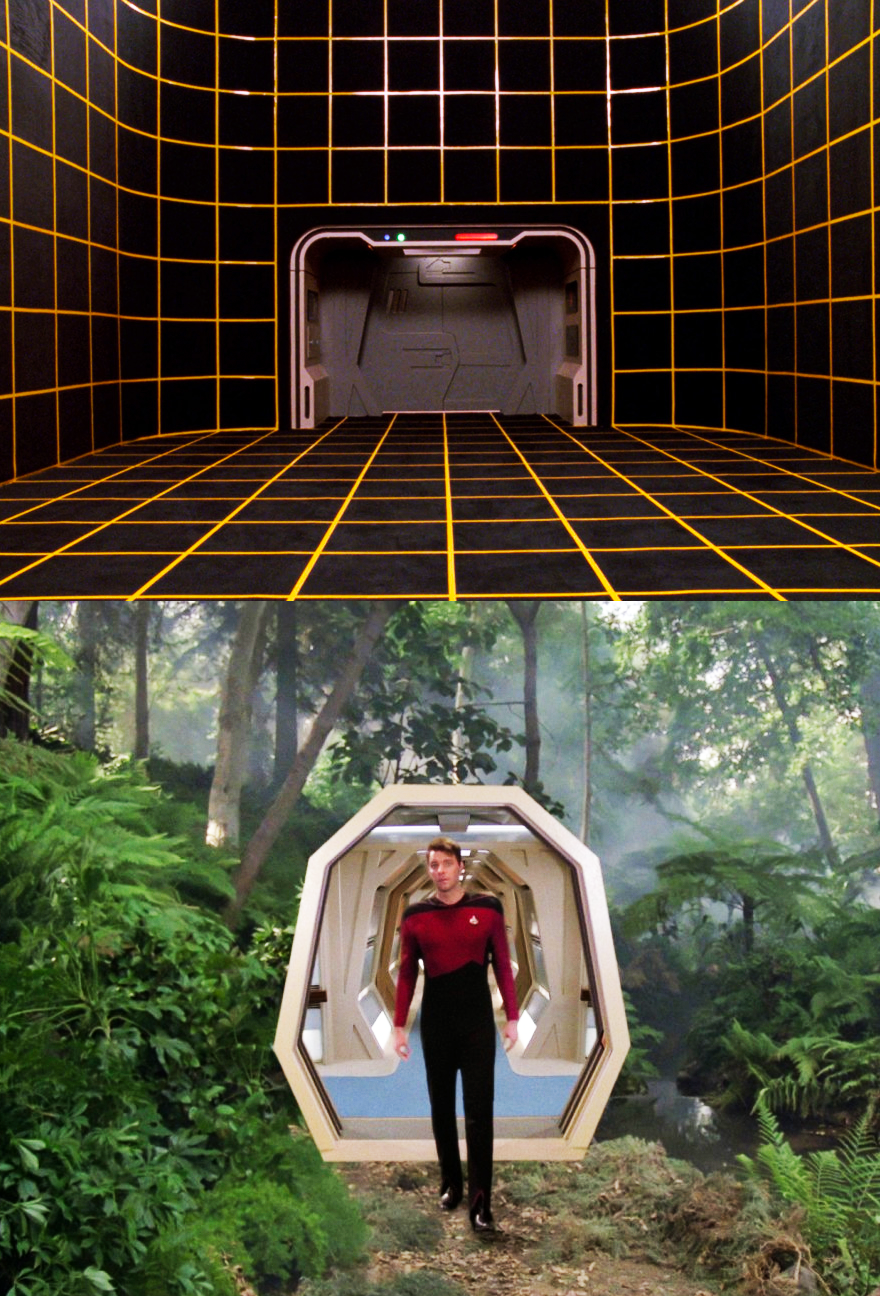}
\caption{The Star Trek Holodeck}
\label{fig:Holodeck}
 \end{figure}
}
combination of different modalities can impact overall perception and quality of an experience.

HCI researchers and designers have begun to explore the viability of integrating multisensory technology into the IOT space~\cite{obrist2016sensing}. For instance, Israr~\cite{israr2014feel}  considered how to enrich storytelling with haptic feedback, Iwata~\cite{iwata2004food} developed a haptic device to simulate biting food, and Spence~\cite{spence2017digitizing} overviewed the state of the art of transferring chemical senses (smell and taste) online. In pursuit of the Holodeck, researchers have also endeavored to integrate different sensory experiences into VR. Lopes~\cite{lopes2015impacto} simulated the physical impact of boxing in VR using electrical muscle simulation, Kiltini~\cite{kilteni2013drumming} explored the role of multisensory and senorimotor feedback in body ownership, and Gerry~\cite{gerry2017paint} superimposed VR on top of physical reality, allowing a novice painter to replicate artist's movements on a canvas while watching them paint in VR. HCI research has likewise attempted to replicate ambient experiences; 
\reversemarginpar
\marginnote{
\begin{figure}
\includegraphics[width=\marginparwidth]{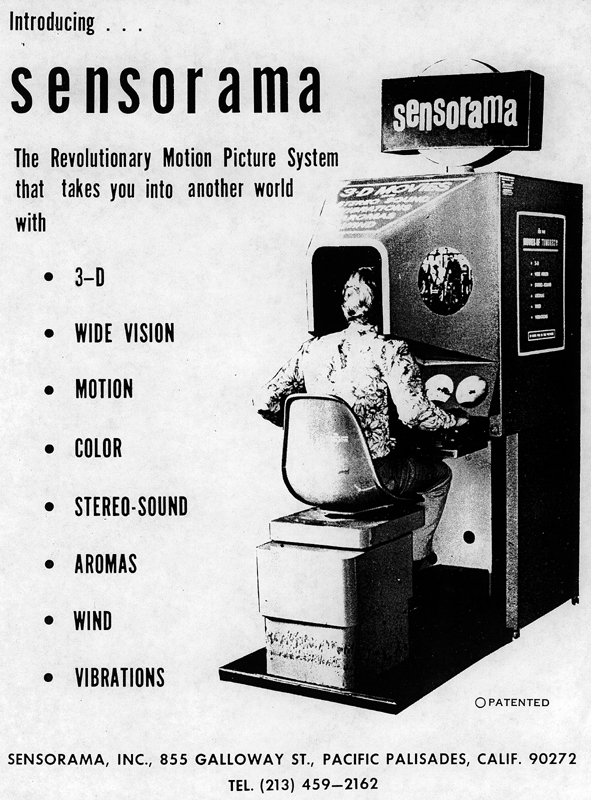}
\caption{Heilig's Sensorama Machine}
\label{fig:Sensorama}
 \end{figure}
} 
Ambioterm~\cite{ranasinghe2017_ambiotherm} simulated environmental conditions in a VR headset, and Martins~\cite{martins2017multisensory_PortWine} conceptualized a  sensory wine tourism experience.

\section{Multisensory Education Design}
Literature suggests that multisensory information can have positive effects on learning. Shams and Seitz~\cite{shams2008benefits} overview the many benefits of multisensory learning on information encoding, storage and retrieval, and note effects on recognition, and cross-modal memory transfer and reinforcement learning. In light of these benefits, HCI has begun to consider the role of multisensory information on engagement and learning; Yannier~\cite{yannier2016adding} found that adding shaking interaction to a virtual learning game helped children enjoy learning physics principles, Covaci and Ghinea~\cite{covaci2018multisensory} found that olfactory information and feedback in an education game engaged students in the task, and Zou~\cite{zou2017_MultisensoryMedia_learnerExperience} found that integrated olfaction, airflow, and haptics stimulti increased learner enjoyment. 
\reversemarginpar
\marginnote{
\begin{figure}
\includegraphics[width=\marginparwidth]{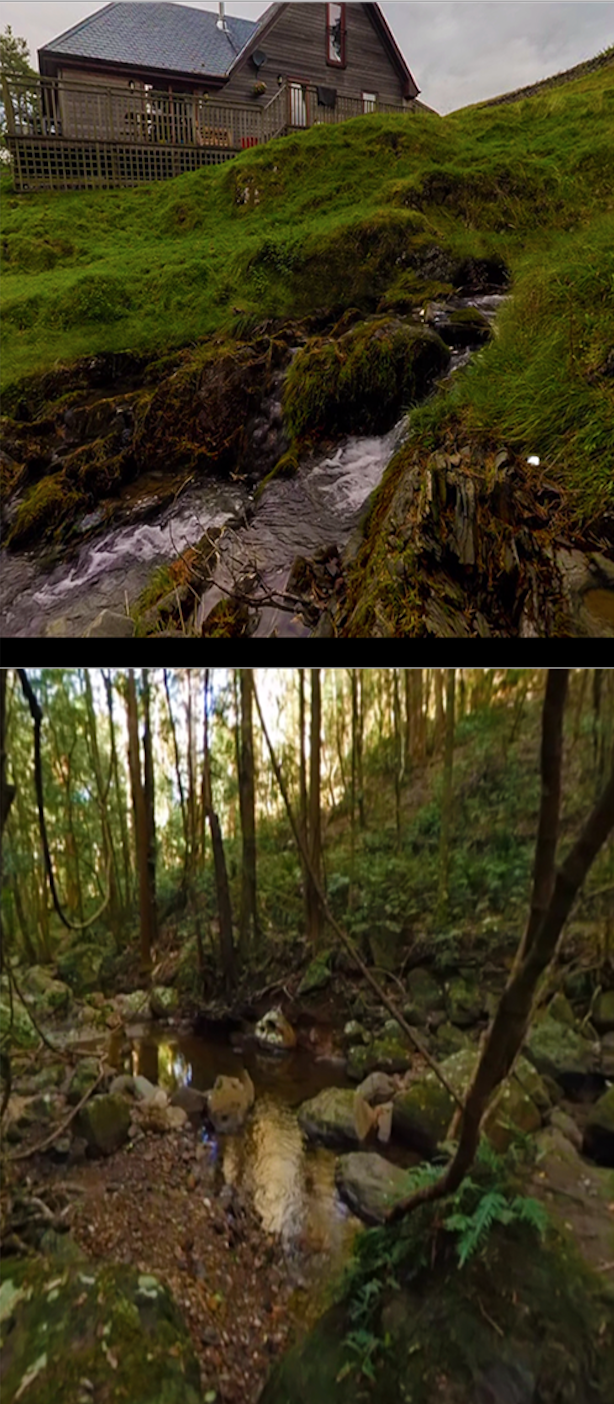}
\caption{Two different streams that learners will experience as 360\textdegree  videos paired with multisensory cues. The top image shows a stream in Scotland, and the bottom shows a stream in the Amazon rainforest.}
\label{fig:VideoStreams}
 \end{figure}
} 
Likewise, multisensory information design has shown to impact judgment and learning tasks. Demate~\cite{dematte2007olfactory} found that olfactory cues could influence people’s judgments of facial attractiveness, Yannier~\cite{yannier2016adding} and Brooks~\cite{brooks1990project} found that haptics helped improve visualization of complex data sets, and Lee~\cite{lee_Spence2008_MultimodalDriverFeedback} found that visual, auditory, and tactile feedback improved virtual racecar drivers' performance. 

\subsection{Multisensory Learning in VR}
Research has also begun conceptualizing the role of multisensory cues on learning in VR. Early works by Psotka~\cite{psotka1995immersive} and Dihn~\cite{dinh1999evaluating} suggest that multisensory cues in VR can reduce conceptual load, create salient memories and emotional experiences, and increase memory and sense of presence for environmental information. As well as creating presence and vivid memories, Dede (1999)~\cite{dede1999multisensory} found that multisensory information helps students understand complex scientific models through experiential metaphors and analogies, which can help displace intuitive misconceptions. In recent work on stream identification tasks, Dede (2017)~\cite{dede2017virtual} suggested that VR could help make topographic characteristics of the watershed more apparent and enhance transfer from virtual settings to the real world; sensory information (e.g. sound, color and turbidity of the water, weather variables, shifts in grass color) could help learners sense pattern changes.

\section[Testing Multisensory VR Pattern Recognition and Knowledge Transfer]{Testing Multisensory VR Pattern\\ Recognition and Knowledge Transfer}

Although research has begun to consider the benefits of mulsemedia in VR, researchers have yet to consider whether training sensory pattern recognition in physical spaces can be simulated, and whether learners can use their experience with simulated multisensory cues to make judgments in physical spaces. My work on StreamBED VR~\cite{striner2016streambed} teaches citizen scientists, volunteers who collaborate with researchers on scientific data collection~\cite{bonney2009citizen}, to make qualitative assessments of local watersheds using the EPA's Rapid Bioassessment Protocol~\cite{barbour1999rapid}. Qualitative tasks are not taught to citizen scientists because of the high cost of teaching background and onsite skills~\cite{Pond:Personal}, so my work considers the viability of VR to train non-experts to make abstract judgments of quality. Being able to train qualitative identification tasks to citizen scientists in VR has the potential to improve data quality and participant retention~\cite{striner2016streambed, WigginsCrowston_DataQuality2011}, and to overcome onsite training costs and challenges.

My earlier work found that experts make stream quality judgments using multiple senses, and that novice learners require a high sense of realism to make accurate judgments. To address learners' need for realism and replicate how experts make assessments, my current work considers the combined effect of audio, olfactory, thermal, and wind cues on pattern recognition. In this study, I plan to compare participants' ability to make observations and recognize patterns while watching 360\textdegree  videos in VR with and without additional multisensory cues; study findings will help develop a better training system, and inform researchers whether ambient environment cues can be virtually recreated. 

\subsection{Study Setup}
During the study, participants will experience 360\textdegree  videos of two streams with similar characteristics and two streams with different characteristics; for instance, figure~\ref{fig:VideoStreams} shows screenshots of streams with different characteristics, a stream from Scotland (top), and the Amazon (bottom).

We are testing the compound effects of several characteristics: video, audio, scent, wind, humidity, and temperature that differ by region. 
\reversemarginpar
\marginnote{
\begin{figure}
\includegraphics[width=\marginparwidth]{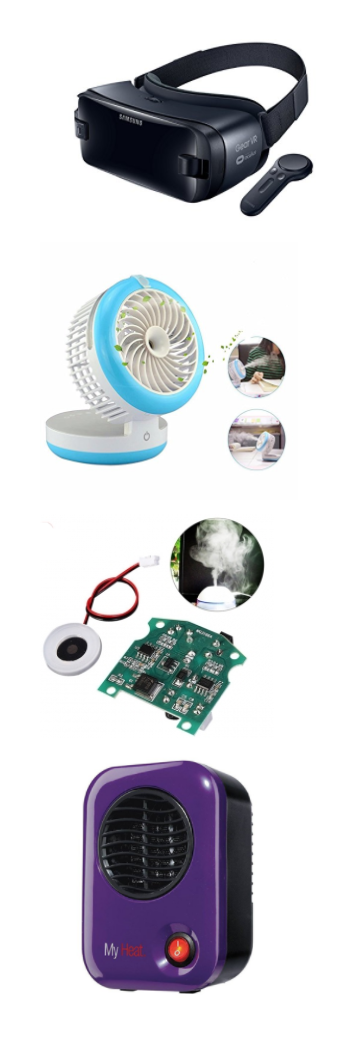}
\caption{Multisensory Setup for Video Study}
\label{fig:MultisensorySetup}
 \end{figure}
} The stream in Scotland would be windy and cool, sound of red deer, water bats and otters, and smell like thistle, wisteria, and limestone. In contrast, the Amazonian stream would have high heat and humidity, sound of cicadas, howler monkeys, and Poison dart frogs, and smell like decomposed wood, soil, and bananas. Participants will experience videos either with or without these multisensory cues. Participants in the multisensory cue condition will experience information through a series of Arduino controlled diffusers, a mister and fan, and a mini desk space heater, shown in figure~\ref{fig:MultisensorySetup}.

\subsection{Challenges}
\textit{Scent Overlap and Removal.} Gallace~\cite{gallace2012multisensory} describe the challenge of olfactory design; it is easy to deliver odors, however creating believable scents, changing and removing odors is problematic. I plan to use manufactured scents, and  will  deliver and extract fans using individual diffuser fans, activated charcoal, and a fume extractor. 

\textit{Multisensory Integration and Cross Modal Stimuli.} Stein~\cite{stein2008_multisensoryIntegration} and others suggest that learners integrate and  synthesize information from cross-modal stimuli. Future work should thus consider how combinations of different cues shape learner mental models.

\textit{Limited Cognitive Resources.} Gallace~\cite{gallace2012multisensory} also describes the 
 limit of cognitive resources for multisensory information processing; incorporating additional senses adds to processing capacity, but decreases the accuracy of judgments for individual variables, resulting in crude judgments of simultaneous things. In order to be effective, multisensory experiences should be "neurally-inspired" by our brain mechanics. While not in the scope the described study, my work plans to consider cognitive limitations when iterating on the StreamBED training system.

\section{Conclusions}
 This paper overviews the state of multisensory interaction design, contends the need to study the effect of mulsemedia on pattern recognition and knowledge transfer in VR, and overviews a study plan to consider multisensory cues on pattern recognition in citizen science stream monitoring training. Understanding how additional sensory information contributes to realism, immersion, and knowledge transfer will help the HCI community design more effective and meaningful IOT systems.

\section{Acknowledgements}
Thanks to Dr. Jennifer Preece for her time and feedback.

\balance
\bibliographystyle{acm-sigchi}
\bibliography{sample}

\end{document}